\documentclass[english,aps,notitlepage,nofootinbib,tightenlines,superscriptaddress, twocolumn]{revtex4-1}
\usepackage[T1]{fontenc}
\usepackage[latin9]{inputenc}
\usepackage{amsmath}
\usepackage{amssymb}
\usepackage{graphicx}

\makeatletter

\pdfoutput=1
\allowdisplaybreaks
\usepackage[colorlinks=true,citecolor=blue,linkcolor=blue]{hyperref}
\usepackage{tabularx}

\makeatother

\usepackage{babel}
\begin{document}
\title{Wash-in leptogenesis after the evaporation of primordial black holes}
\author{Kai Schmitz}
\email{kai.schmitz@uni-muenster.de}
\affiliation{Institute for Theoretical Physics, University of M\"unster, 48149 M\"unster, Germany}
\author{Xun-Jie Xu}
\email{xuxj@ihep.ac.cn}
\affiliation{Institute of High Energy Physics, Chinese Academy of Sciences, Beijing 100049, China}
\date{\today }
\hfill MS-TP-23-46
\begin{abstract}
Wash-in leptogenesis is a powerful mechanism to generate the baryon
asymmetry of the Universe that treats right-handed-neutrino interactions
on the same footing as electroweak sphaleron processes: as mere spectator
processes acting on the background of chemical potentials in the Standard
Model plasma. Successful wash-in leptogenesis requires this chemical
background to be $CP$-violating, which can be achieved by violating
any of the more than ten global charges that are conserved in the
Standard Model at very high temperatures. In this paper, we demonstrate
that the primordial charge asymmetries required for wash-in leptogenesis
can be readily produced by evaporating primordial black holes (PBHs).
Our argument is based on the fact that the Hawking radiation emitted
by PBHs contains more or less any state in the particle spectrum.
Therefore, if heavy states with $CP$-violating decays are present in
the ultraviolet, PBH evaporation will unavoidably lead to the production
of these states. We illustrate this scenario by means of a simple
toy model where PBH evaporation leads to the production of heavy particles
that we call asymmetrons and whose decay results in a primordial charge
asymmetry for right-handed electrons, which in turn sets the initial
conditions for wash-in leptogenesis. We focus on the parameter region
where the decay of the initial thermal asymmetron abundance occurs
long before PBH evaporation and only results in a negligible primordial
charge asymmetry. PBH evaporation at later times then serves as a
mechanism to resurrect the asymmetron abundance and ensure the successful
generation of the baryon asymmetry after all. We conclude that PBHs
can act as asymmetry-producing machines that grant access to whatever
$CP$-violating physics may be present in the ultraviolet, rekindling
it at lower energies where it can be reprocessed into a baryon asymmetry
by right-handed neutrinos.
\end{abstract}
\maketitle

\section{Introduction }

The Standard Model (SM) fails to provide an explanation for the observed baryon asymmetry of the Universe,
usually expressed in terms of the baryon-to-photon ratio,
$\eta_B^0 = n_b^0/n_\gamma^0 \simeq 6.1 \times 10^{-10}$~\cite{Planck:2018vyg,ParticleDataGroup:2022pth}, which hence provides clear 
evidence for new physics. A popular scenario for the dynamical generation of
the baryon asymmetry in the early Universe consists of thermal leptogenesis~\cite{Fukugita:1986hr}, which, in its 
simplest form, is based on the charge--parity ($CP$) and baryon-minus-lepton number ($B\!-\!L$) violating 
out-of-equilibrium decays of heavy right-handed neutrinos (RHNs). These decays first generate a primordial
$B\!-\!L$ asymmetry, which is subsequently partially converted to a baryon asymmetry by
the chemical transport in the SM plasma, including electroweak sphalerons~\cite{Kuzmin:1985mm}.
Baryogenesis via leptogenesis is an attractive scenario that links the
generation of the matter--antimatter asymmetry in the early Universe to new physics in the neutrino
sector that may be within the reach of laboratory experiments~\cite{Buchmuller:2005eh,Chun:2017spz,Bodeker:2020ghk}.
Thermal leptogenesis in its standard
formulation, however, comes with several restrictions that render its experimental
exploration very challenging. For instance, in order to achieve a sufficient amount
of  $CP$ violation in RHN decays, the mass of the lightest RHN mass eigenstate is bounded
from below, $M_1 \gtrsim 10^{9}\,\textrm{GeV}$~\cite{Davidson:2002qv,Buchmuller:2002rq}.
Similarly, the Yukawa interactions that couple RHNs to SM lepton--Higgs pairs must not
be too strong, since otherwise wash-out effects will erase any previously
generated asymmetry and hence spoil the success of leptogenesis.

There are numerous proposals for alternative leptogenesis scenarios at lower energies, notably
resonant leptogenesis~\cite{Pilaftsis:2003gt,Pilaftsis:2005rv,Dev:2017wwc} and leptogenesis
via RHN oscillations~\cite{Akhmedov:1998qx,Drewes:2017zyw} (see Refs.~\cite{Klaric:2020phc,Klaric:2021cpi}
for recent work on the relation between these two models). 
These scenarios partially rely on special choices of parameter values (e.g., a highly
degenerate RHN mass spectrum) and  continue to exploit the $CP$ violation in
the RHN sector in order to generate a $B\!-\!L$ asymmetry. By contrast, the recently
proposed mechanism of wash-in leptogenesis~\cite{Domcke:2020quw,Domcke:2022kfs}
follows a different approach; it dispenses with $CP$ violation in the RHN sector
altogether and separates the energy scales of $CP$ violation and $B\!-\!L$ violation. 

Wash-in leptogenesis generalizes thermal leptogenesis to nontrivial chemical background 
configurations in the early Universe and treats RHN interactions on the same footing
as electroweak sphalerons: as mere spectator processes that reprocess 
the chemical potentials of the SM particle species in the thermal bath. The action 
of the electroweak sphalerons on the chemical composition of the SM plasma then
results in the usual violation of baryon-plus-lepton number $B\!+\!L$, while the
action of the RHN interactions results in the violation
of $B\!-\!L$. Together, these two effects are sufficient for the generation of a
primordial baryon asymmetry.

$CP$ violation in the RHN sector is irrelevant for
wash-in leptogenesis. Instead, wash-in leptogenesis is based on the idea that new 
$CP$-violating dynamics at high energies are responsible for the generation of 
primordial charge asymmetries, which are then reshuffled in a $B\!-\!L$-violating
fashion by RHN interactions at low energies. In this sense, the idea of wash-in leptogenesis
denotes a general mechanism that can be used as a building block in a complete model
of baryogenesis; but it does not represent a complete baryogenesis scenario on its own.
In other words, wash-in leptogenesis requires an ultraviolet (UV) completion that
provides an explanation for the origin of the nontrivial chemical background that 
the RHN interactions are supposed to act on\,---\,a mechanism for chargegenesis in the language
of Refs.~\cite{Domcke:2022kfs}. An interesting and viable scenario
for such a UV completion of wash-in leptogenesis is axion inflation~\cite{Anber:2015yca,Jimenez:2017cdr,Domcke:2019mnd,Domcke:2022kfs},
i.e., models of inflation where inflation is driven by a pseudoscalar axion field that spontaneously
breaks $CP$ invariance by means of its nonzero and time-dependent background field value. If the axion field
couples to the SM hypercharge gauge field, the $CP$ violation induced by the dynamics of the rolling
axion field is reflected in the dual production of (A) maximally helical hypermagnetic
fields and (B) fermionic charge asymmetries in accordance with the chiral anomalies of the SM fermion
currents. These charge asymmetries then provide the necessary initial conditions
for wash-in leptogenesis at lower temperatures.

Wash-in leptogenesis after axion inflation is, however, only one example
scenario among countless other possibilities. In fact, wash-in leptogenesis
only requires that at least one of various global charges that are conserved
in the SM at high temperatures be violated. Depending on the temperature scale
of wash-in leptogenesis, there are up to eleven suitable charges whose violation
can set the right initial conditions for wash-in leptogenesis, 
\begin{align}
\label{eq:charges}
& q_e \,,\quad q_{2 B_1 - B_2 - B_3} \,,\quad q_{u-d} \,,\quad q_{d-s} \,,\quad q_{B_1 - B_2} \,, \\ 
& q_\mu \,,\quad q_{u-c} \,,\quad q_\tau \,,\quad q_{d-b} \,,\quad q_B \,,\quad q_u \,, \nonumber
\end{align}
see Refs.~\cite{Domcke:2020quw,Domcke:2022kfs} for more details.
This characteristic feature of wash-in leptogenesis renders it
an extremely flexible and general mechanism that can lead to the successful
generation of the baryon asymmetry in a plethora of models. The purpose of this paper
is to highlight one such class of models, which has not yet been discussed in
the literature: wash-in leptogenesis after the evaporation of
primordial black holes (PBHs)~\cite{Escriva:2022duf}. 

The key observation behind this scenario is that the Hawking radiation emitted by evaporating
PBHs contains particles across the whole particle spectrum, including heavy states
that are potentially never produced thermally or that already freeze out at very
high temperatures. In order to generate the baryon asymmertry of the Universe, it is
then enough to assume that there is at least one state in the particle spectrum whose
interactions result in a sufficient amount of $CP$ violation, such that it can generate the
right initial conditions for wash-in leptogenesis. We shall refer to this
particle as the asymmetron. Given a population of very light PBHs in the early Universe, the evaporation
of these PBHs will unavoidably result in the production of asymmetrons, whose interactions then yield
the charge asymmetries required for wash-in leptogenesis. In our scenario, PBH evaporation 
thus plays the role of the chargegenesis mechanism that serves as the UV completion of
wash-in leptogenesis.

Our scenario is related to baryogenesis~\cite{Hawking:1974rv,Carr:1976zz,Barrow:1990he,Baumann:2007yr,Morrison:2018xla,Hooper:2020otu,Gehrman:2022imk}
and leptogenesis~\cite{Perez-Gonzalez:2020vnz,Datta:2020bht,JyotiDas:2021shi,Bernal:2022pue,Calabrese:2023key,Khan:2023myt}
models that rely on PBH evaporation in order to produce an
abundance of particles whose decays either violate baryon number $B$ or lepton number $L$ (i.e., RHNs in
the latter case). Meanwhile, our scenario is more general than these earlier proposals. In our case,
we merely need to assume that the $CP$-violating asymmetron interactions produce any of the eleven global charges
in Eq.~\eqref{eq:charges}, which are conserved in the SM at high temperatures; we do not need
to identify the asymmetron with, say, a RHN and we do not need to assume that the asymmetry
produced in asymmetron interactions corresponds to $B$ or $L$ right away. Instead, PBHs play the
role of generic asymmetry-producing machines in our scenario that provide a portal to whatever
$CP$-violating physics may be present at high energies. PBH evaporation grants access 
to these $CP$-violating dynamics and leads to a nontrivial chemical equilibrium in one form
or another, which is all it takes to realize wash-in leptogenesis.

In this paper, we will illustrate the basic idea of wash-in leptogenesis after PBH evaporation
by means of a simple toy model, in which PBH evaporation gives rise to an abundance of asymmetrons whose decays 
result in a primordial asymmetry between right-handed electrons and left-handed positrons.
We expect that this toy model can be easily generalized to more complex scenarios, involving other
primordial charge asymmetries or even combinations thereof. The rest of the paper is organized
as follows: We begin in Sec.~\ref{sec:basic} by introducing our toy model. Then, in Sec.~\ref{sec:ana},
we provide a few analytical estimates, before we study the full system of Boltzmann equations in
Sec.~\ref{sec:boltzmann}. In Sec.~\ref{sec:Full-scan}, we perform a systematic scan of parameter
space, before we finally conclude in Sec.~\ref{sec:con}.

\section{ A simple scenario \label{sec:basic}}

The idea of wash-in leptogenesis can be illustrated by considering
a particularly simple scenario in the early universe when the temperature
is above ${\cal O}(100)$ TeV, at which the electron Yukawa interaction
is not equilibrated. 

Therefore, any primordial asymmetry of the right-handed electron ($e_{R}$)
produced at higher temperatures can be retained until ${\cal O}(100)$
TeV.  If the wash-out effect due to for example  RHN interactions
is strong,   the $B-L$ asymmetry is then mainly determined by wash-in
leptogenesis, which generates~\cite{Domcke:2020quw}
\begin{equation}
q_{B-L}=-\frac{3}{10}q_{e}\thinspace,\label{eq:washin}
\end{equation}
where $q_{e}=n_{e_{R}}-n_{\overline{e_{R}}}$ with $n_{e_{R}}$ and
$n_{\overline{e_{R}}}$ the number densities of $e_{R}$ and $\overline{e_{R}}$,
and $q_{B-L}=\sum_{f}Q_{B-L}^{(f)}(n_{f}-n_{\overline{f}})$ with
$n_{f}$ and $Q_{B-L}^{f}$ the number density and the $B-L$ charge
of particle $f$. 

Equation~\eqref{eq:washin} implies that if there
is a mechanism at high temperatures giving rise to nonzero $q_{e}$, there should be $B-L$
asymmetry at low temperatures. To generate nonzero $q_{e}$, we consider
PBHs which can emit particles via Hawking radiation quite generically,
independent of the particle interactions. In particular, they can
emit extremely heavy particles that could otherwise not be produced
in the early universe due to the finite reheating temperature. For
a given PBH of mass $m_{{\rm BH}}$, the rate of emission is
\begin{align}
\frac{d^{2}N_{i}}{dtdE} & \approx\frac{g_{i}}{2\pi}\cdot\frac{v_{i}}{\exp(E/T_{{\rm BH}})\pm1}\thinspace,\label{eq:-39}\\
T_{{\rm BH}} & =\frac{m_{{\rm pl}}^{2}}{8\pi m_{{\rm BH}}}\thinspace,\label{eq:-40}
\end{align}
where $d^{2}N_{i}/dtdE$ denotes the number of particle $i$ emitted
within the differential time $dt$ and the differential energy $dE$;
$g_{i}$ and $v_{i}$ are the multiplicity and graybody factor of
particle $i$;  and $m_{{\rm pl}}=1.22\times10^{19}$ GeV is the Planck
mass. The ``$\pm$'' sign takes $+$ or $-$ for fermions and bosons
respectively.  

The Hawking radiation of a PBH in general emits particles and antiparticles
equally. To generate the asymmetry of $e_{R}$, we introduce a heavy
particle $X$ that can decay asymmetrically to $e_{R}$ and $\overline{e_{R}}$.
For this reason, we dub it ``the asymmetron''. The asymmetron can
be either a Majorana fermion or a real scalar, as it needs to be its
own antiparticle. We assume that it is a Majorana fermion with the
following interaction: 
\begin{equation}
{\cal L}\supset yXe_{R}S\thinspace,\label{eq:-1}
\end{equation}
where $S$ is a scalar with the opposite SM charges of $e_{R}$, i.e.,
 $e_{R}$ and $S$ possess hypercharges $Y_{e_{R}}=-1$ and $Y_{S}=+1$.
As an asymmetron, $X$ needs to decay asymmetrically. Hence we introduce
the parameter of asymmetric decay:
\begin{equation}
\epsilon\equiv\frac{\Gamma_{X\to e_{R}S}-\Gamma_{X\to\overline{e_{R}}S^{*}}}{\Gamma_{X\to e_{R}S}+\Gamma_{X\to\overline{e_{R}}S^{*}}}\thinspace,\label{eq:-2}
\end{equation}
where $\Gamma$ denotes the decay width of the process indicated in
the subscript. The asymmetric decay can be attained by for example
assigning a flavor structure to Eq.~\eqref{eq:-1} with $CP$ phases,
similar to the asymmetric decay of right-handed neutrinos. In general,
we expect $\epsilon\lesssim y^{2}/(8\pi)$ if all Yukawa couplings
in the flavorful generalization are ${\cal O}(y)$. Much smaller $\epsilon$
is possible by tuning down the $CP$ phases in the flavor structure.

\section{Low temperature, High mass\protect \\
---\,an analytical estimate\label{sec:ana}  }

One of the most prominent features of PBHs is that they can emit heavy
particles with masses much higher than the environmental temperature.
For heavy asymmetrons produced at a relatively late epoch of the universe
when $T\ll m_{X}$,  their decay is highly out-of-equilibrium,  which
is important for the produced asymmetry to evade thermal washout.
Due to this feature, within a certain range of the parameter space,
the asymmetry $q_{e}$ would predominantly depend on the yield of
$X$ from PBH evaporation.

This is the scenario we will concentrate on in what follows.  To
make the  discussion more concrete, let us set a benchmark: 
\begin{align}
 & m_{{\rm BH0}}=10\ {\rm g}\thinspace,\ m_{X}=10^{12}\ {\rm GeV}\thinspace,\label{eq:b1}\\
 & y=0.5\thinspace,\ \beta=10^{-6}\thinspace,\ \epsilon=-0.068\times\frac{y^{2}}{8\pi}\thinspace,\label{eq:b2}
\end{align}
where $m_{{\rm BH0}}$ is the initial mass of PBHs, and $\beta\equiv\rho_{{\rm BH0}}/\rho_{{\rm tot}}$
with $\rho_{{\rm BH0}}$ the initial energy density of PBHs and $\rho_{{\rm tot}}$
the total energy density of the thermal bath. For simplicity, we assume
that all PBHs have the same mass. 
The numeric factor for $\epsilon$ is set at 0.068 such that the produced asymmetry meets the observed value, as we will see later.

\begin{figure}
\centering

\includegraphics[width=0.99\columnwidth]{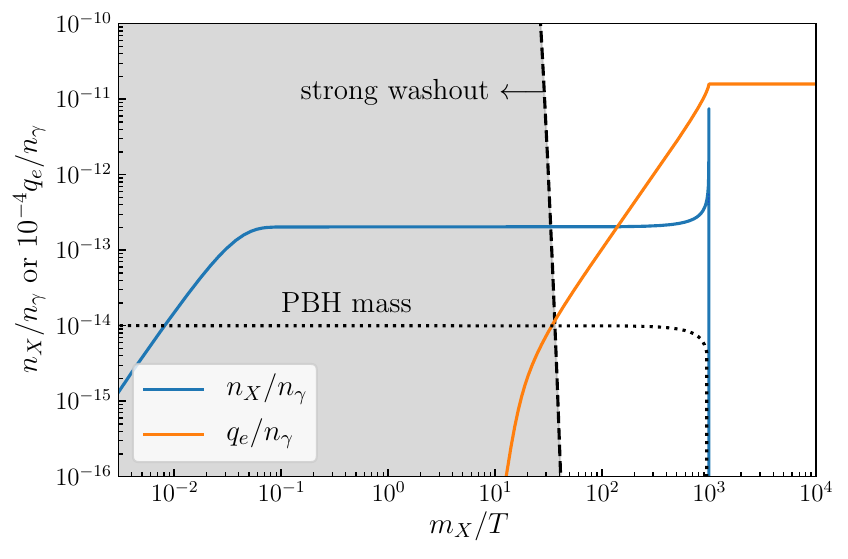}

\caption{Evolution of the asymmetron number density, $n_{X}$, and the electron
asymmetry, $q_{e}$, for the benchmark in Eqs.~\eqref{eq:b1}-\eqref{eq:b2}.
Note that here $n_X$ and $q_e$ denote only the contributions of asymmetrons produced from PBHs, while the thermal asymmetron abundance is also present (the dashed line), though it leads to negligible asymmetry due to strong washout. 
The dotted curve represents the  value of $10^{-14}m_{\text{BH}}/m_{\text{BH}0}$. 
\label{fig:bench}}
\end{figure}

Figure~\ref{fig:bench} shows the evolution of the number density of $X$ produced from PBHs
for this benchmark. The result is obtained from our numerical calculation
in Sec.~\ref{sec:boltzmann}. Analytically, we can also understand
the behavior of the curve well and provide a quite accurate estimate
of the result.  

According to  Eq.~\eqref{eq:-40}, the initial temperature of the
PBHs in Eq.~\eqref{eq:b1} is $T_{\text{BH}0}\approx1.06\times10^{12}$
GeV, which is comparable to $m_{X}$. Hence they can efficiently emit
$X$ particles. As the PBHs keep evaporating, the PBH temperature
and hence the emission rate keeps increasing. Eventually the PBHs
will fully evaporates with the lifetime given by\footnote{See Appendix \ref{sec:Basic-PBH} for a review of the basic aspects
of PBH.}
\begin{equation}
\tau_{{\rm BH}}=\frac{m_{{\rm BH}0}^{3}}{3g_{{\rm BH}}m_{\text{pl}}^{4}}\thinspace,\label{eq:-3}
\end{equation}
where $g_{{\rm BH}}\approx7.5\times10^{-3}$ if all SM degrees of
freedom are included. By  comparing $\tau_{{\rm BH}}$ with the
time of the universe, $t\approx1/(2H)$ where $H$ is the Hubble parameter,
we obtain that the PBHs should fully evaporate when the universe cools
down to
\begin{equation}
T_{{\rm eva}}\approx10^{9}\ \text{GeV}\times\left(\frac{10\ {\rm g}}{m_{{\rm BH0}}}\right)^{3/2}\thinspace.\label{eq:-6}
\end{equation}

The formation of PBHs is related to cosmological perturbations at
the horizon scale ($H^{-1}$), leading to $m_{{\rm BH0}}\propto4\pi H^{-3}\rho_{{\rm tot}}/3\propto T^{-6}\cdot T^{4}\propto T^{-2}$.
Hence smaller PBHs are formed at higher temperatures.  More specifically,
the temperature of the universe at which the PBHs with initial mass $m_{{\rm BH0}}$
are formed is given by \cite{Bernal:2022pue}
\begin{equation}
T_{0}\approx1.35\times10^{15}\ \text{GeV}\times\left(\frac{10\ {\rm g}}{m_{{\rm BH0}}}\right)^{1/2}\thinspace.\label{eq:-8}
\end{equation}
  So the entire PBH life for the given benchmark spans from $T_{0}\sim10^{15}$
GeV to $T_{{\rm eva}}\sim10^{9}$ GeV. 

Despite that these PBHs keep evaporating during the period from $10^{15}$
to $10^{9}$ GeV,   their masses stay almost static until the very
last moment shortly before $T_{\text{eva}}$---see the dotted curve
in Fig.~\ref{fig:bench}. Consequently, the emission rate of $X$
is almost a constant at $T\gg T_{{\rm eva}}$. This leads to the flatness
of the blue curve in Fig.~\ref{fig:bench} for $m_{X}/T\in[4\times10^{-2},\ 7\times10^{2}]$. 

The flat part of the curve can be estimated by assuming  the balance
between the production of $X$ from PBHs and the depletion
due to $X$ decay:
\begin{equation}
n_{\text{BH}}\Gamma_{\text{BH}\to X}=n_{X}\Gamma_{X}\frac{m_{X}}{2T_{\text{BH}}}\thinspace,\label{eq:-9}
\end{equation}
where $n_{\text{BH}}$ denotes the number density of PBHs, 
$\Gamma_{\text{BH}\to X}$ 
denotes the emission rate of $X$ per PBH, and $\Gamma_{X}=y^{2}m_{X}/(16\pi)$
is the decay rate of $X$ at rest. The PBH number density $n_{\rm BH}$ is determined by rescaling its initial value, $n_{{\rm BH}}=n_{{\rm BH0}}\,a_0^3/a^3$ with $n_{{\rm BH0}}=\beta {\rho_{\rm tot}}/{m_{\text{BH0}}}$. 
For $\Gamma_{\text{BH}\to X}$, we take $\Gamma_{\text{BH}\to X}\approx0.012\,T_{{\rm BH}}$---see Appendix \ref{sec:Basic-PBH} for details. 
The $m_{X}/(2T_{{\rm BH}})$
factor accounts for the time dilution effect of relativistic $X$
decay, since the average value of $m_X/E$ is around $0.46\, m_{X}/T_{{\rm BH}}$.

Solving Eq.~\eqref{eq:-9} with respect to $n_{X}$, we obtain
\begin{equation}
\frac{n_{X}}{n_{\gamma}}\approx0.28\frac{\beta m_{\text{pl}}^{4}T_{0}}{m_{{\rm BH0}}^{3}m_{X}^{2}y^{2}}\thinspace,\label{eq:-14}
\end{equation}
where $n_{\gamma}=2\zeta(3)T^{3}/\pi^{2}$ is the photon number density.
For the benchmark values in Eqs.~\eqref{eq:b1}-\eqref{eq:b2}, 
Eq.~\eqref{eq:-14}
gives $n_{X}/n_{\gamma}\approx 1.9\times10^{-13}$, 
agreeing well with
the flat part of the blue curve in Fig.~\ref{fig:bench}. 

At high temperatures ($m_{X}/T\lesssim30$), the number density $n_{X}$
in Eq.~\eqref{eq:-14} is much lower than its thermal equilibrium
value, $n_{X}^{{\rm eq}}$,  which is plotted in Fig.~\ref{fig:bench}
as the black dashed curve. In this regime, any asymmetry produced
via $X$ decay can be very efficiently washed out by thermal processes
($e_{R}S\leftrightarrow X\leftrightarrow\overline{e_{R}}S^{*}$).
Therefore, we regard the gray region encompassed by the dashed curve
as the strong washout region. Within this region, the asymmetry $q_{e}$
generated from $X$ decay is strongly suppressed. Only when $n_{X}$
stretches out of the gray region, the generated asymmetry can be effectively
retained, as can be seen from the orange curve in Fig.~\ref{fig:bench}. 

Therefore, the final contribution of PBH evaporation to the asymmetry
can be estimated by
\begin{equation}
q_{e}\approx n_{{\rm BH}}\epsilon\int_{t_{\text{eq}}}^{t_{{\rm eva}}}\Gamma_{\text{BH}\to X}dt\thinspace,\label{eq:-15}
\end{equation}
where $t_{{\rm eq}}$ and $t_{{\rm eva}}$ denote the moments 
when the blue curve crosses the dashed curve and when the PBHs fully
evaporate, respectively. The result is  
\begin{equation}
\frac{q_{e}}{n_{\gamma}}\approx0.0073\epsilon\beta\frac{m_{\text{pl}}^{2}T_{0}}{m_{{\rm BH0}}\left(T_{{\rm BH}}^{{\rm eq}}\right)^{2}}\thinspace,\label{eq:-16}
\end{equation}
where $T_{{\rm BH}}^{{\rm eq}}$ denotes the PBH temperature at $t=t_{{\rm eq}}$.
Taking the benchmark values in Eqs.~\eqref{eq:b1}-\eqref{eq:b2},
we obtain  $q_{e}/n_{\gamma}\approx -1.6\times10^{-7}$, agreeing well
with the orange curve in Fig.~\ref{fig:bench}. 

To account for the observed baryon asymmetry, $\eta_{B}\equiv(n_{B}-n_{\overline{B}})/s\approx8.6\times10^{-11}$
where $s\approx7.04n_{\gamma}$ for the present universe or $192.2n_{\gamma}$
when all SM species are relativistic, wash-in leptogenesis requires
$q_{e}/n_{\gamma}=192.2\eta_{B}\times(-10/3)\times(79/28)=-1.6\times10^{-7}$.
Therefore, the above benchmark produces exactly the observed baryon asymmetry.

If the asymmetron mass is well above the initial temperature of the
PBHs, $m_{X}\gg T_{{\rm BH}0}$, then we also need to take into account
that the emission of $X$ becomes significant only when the PBH temperature
increases to $T_{{\rm BH}}\gtrsim{\cal O}(m_{X})$. In practice, we
find that it is a good approximation to replace $T_{{\rm BH}}^{{\rm eq}}\to\max(T_{{\rm BH}}^{{\rm eq}},\ m_{X}/3.6)$
in Eq.~\eqref{eq:-16} to account for the heavy mass suppression. 

Here we would like to compare our calculation to a similar calculation
in Ref.~\cite{Bernal:2022pue}, which also analytically estimated
the production of heavy generic particles ($X$) from PBHs and the
associated asymmetry. In Ref.~\cite{Bernal:2022pue}, the authors
assume that all $X$ particles decay only after the PBHs fully evaporate,
while here we assume that the decay is rapid, causing a balance between
the production and depletion of $X$. The two different scenarios
correspond to small and large $y$. While we do find that there is
a limited space (see Sec.~\ref{sec:Full-scan}) for the former scenario
to generate the required asymmetry predominantly from PBHs, we mainly
focus on the latter as the viable parameter space is significantly
larger. In fact, our Eq.~\eqref{eq:-16} can be readily applied to
the scenario with small $y$ and long-lived $X$, because Eq.~\eqref{eq:-15}
does not involve any assumptions on the evolution of $n_{X}$. As
we have checked, Eq.~\eqref{eq:-16} approximately reproduces the
analytical result of Ref.~\cite{Bernal:2022pue} in the weak washout
regime.

Finally, let us estimate a potentially relevant process that could
significantly wash out the produced asymmetry at a relatively late
epoch. At $T\ll m_{X}$, the two-to-two scattering $e_{R}S\leftrightarrow\overline{e_{R}}S^{*}$
becomes the dominant process of washout, with the reaction rate 
\begin{equation}
\Gamma_{e_{R}S\leftrightarrow\overline{e_{R}}S^{*}}\approx\frac{y^{4}T^{3}}{32\pi^{3}m_{X}^{2}}\thinspace.\label{eq:G-2-2}
\end{equation}
By comparing $\Gamma_{e_{R}S\leftrightarrow\overline{e_{R}}S^{*}}$
with the Hubble expansion rate, we see that the scattering washout
becomes ineffective ($\Gamma_{e_{R}S\leftrightarrow\overline{e_{R}}S^{*}}\lesssim H$)
when $T$ is below 
\begin{equation}
T_{{\rm dec}}\approx\frac{1.4\times10^{9}{\rm GeV}}{y^{4}}\left(\frac{m_{X}}{10^{12}{\rm GeV}}\right)^{2}.\label{eq:Tdec}
\end{equation}
If $T_{{\rm dec}}$ is higher than $T_{{\rm eva}}$ in Eq.~\eqref{eq:-6},
then most of the $X$ particles produced from PBHs decay after the
two-to-two scattering has decoupled. In this case, the washout via
scattering can be neglected. The condition $T_{{\rm dec}}\gtrsim T_{{\rm eva}}$
can be formulated as a bound on $y$:
\begin{equation}
y\lesssim1.1\times\left(\frac{m_{X}}{10^{12}{\rm GeV}}\right)^{1/2}\left(\frac{m_{{\rm BH0}}}{10\ {\rm g}}\right)^{3/8}.\label{eq:y-bound}
\end{equation}
Obviously, the benchmark value of $y$ in Eq.~\eqref{eq:b2} meets
this condition, implying that it is safe to neglect the washout via
scattering for the benchmark. 

In what follows, we only consider that $y$ is below the boundary
given by Eq.~\eqref{eq:y-bound}. Above the boundary, the two-to-two
scattering washout becomes significant, which likely spoils the success
of leptogenesis. We, however, leave a more detailed analysis of this
parameter regime for future work.

\section{The Boltzmann approach \label{sec:boltzmann}}

\begin{figure*}
\centering

\includegraphics[width=0.32\textwidth]{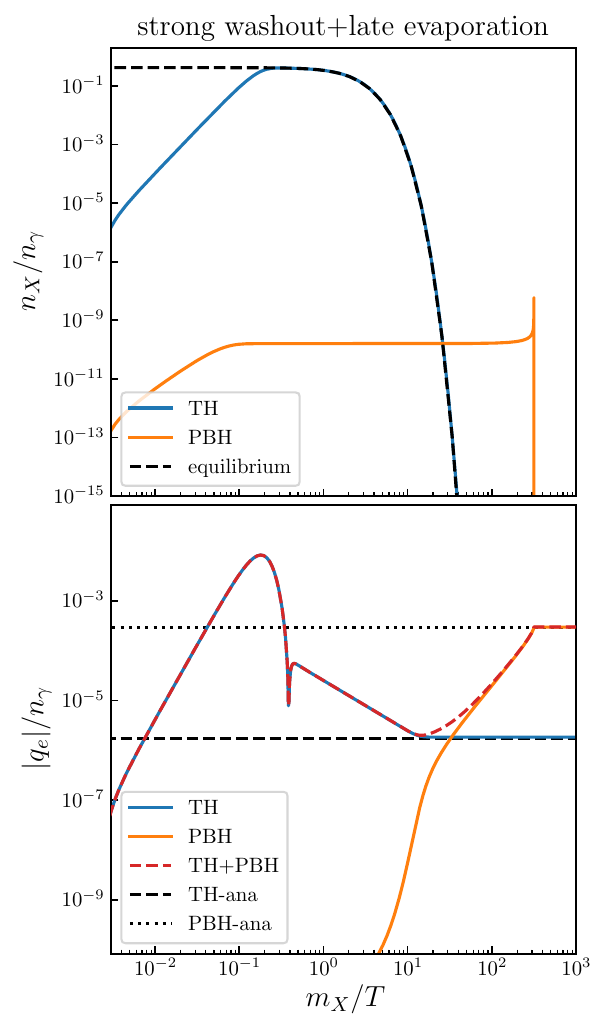}\includegraphics[width=0.32\textwidth]{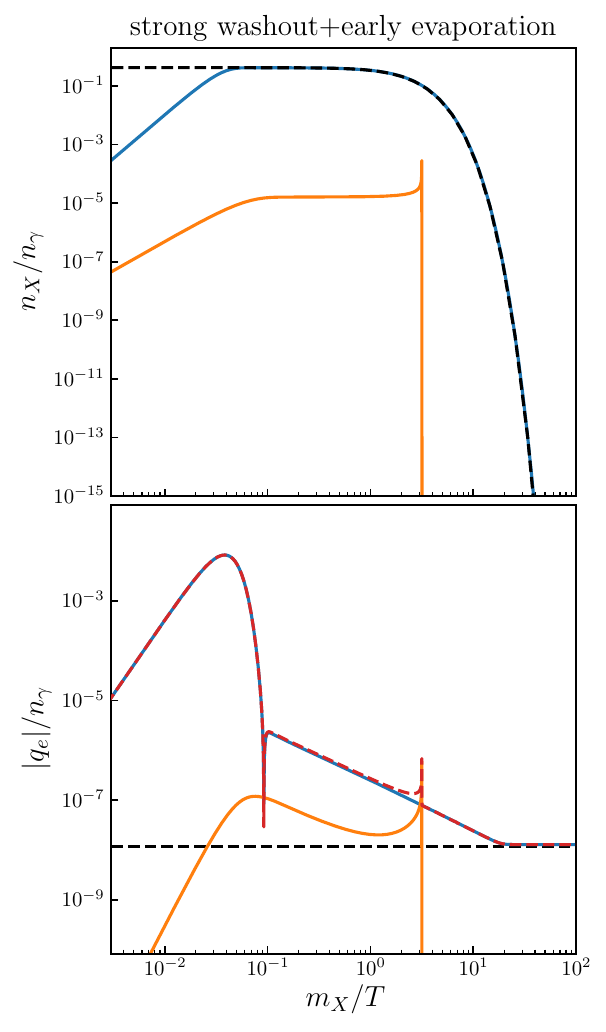}\includegraphics[width=0.32\textwidth]{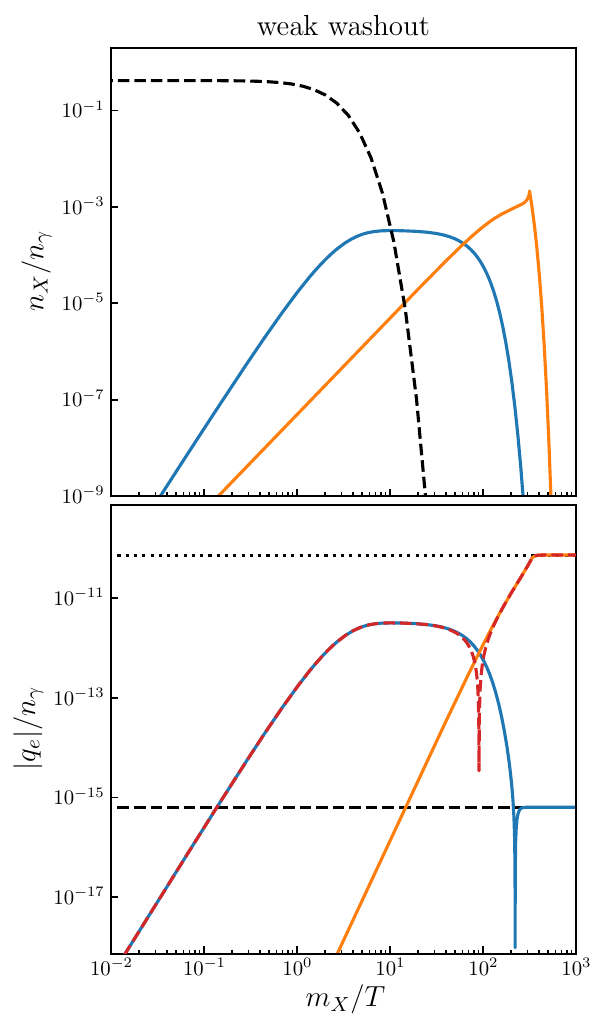}

\caption{Numerical solutions of the Boltzmann equations. From left to right
panels, we select three three representative examples to demonstrate
i) strong washout with late PBH evaporation; ii) strong washout with
early PBH evaporation; iii) weak washout. The labels ``TH'' and
``PBH'' denote thermal and PBH contributions, and ``TH+PBH'' denotes
their sum. The black dotted and dashed lines in the lower panels are
analytical estimates given by Eqs.~\eqref{eq:-16} and \eqref{eq:-25}.
 \label{fig:boltzmann}}
\end{figure*}

\begin{figure*}
	\centering
	
	\includegraphics[width=0.49\textwidth]{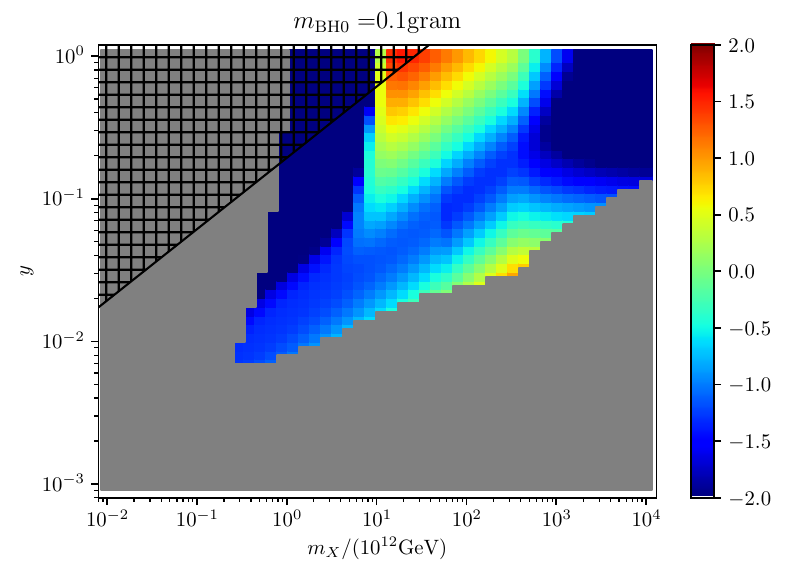}\includegraphics[width=0.49\textwidth]{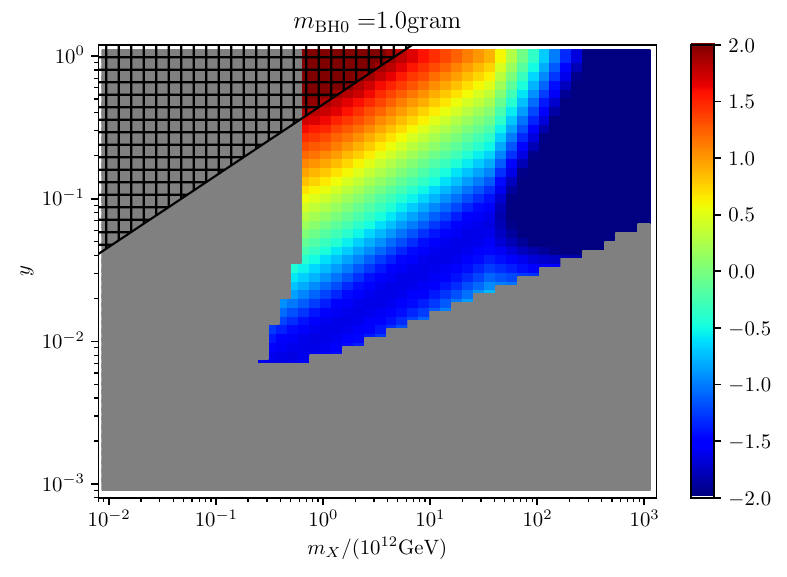}
	
	\includegraphics[width=0.49\textwidth]{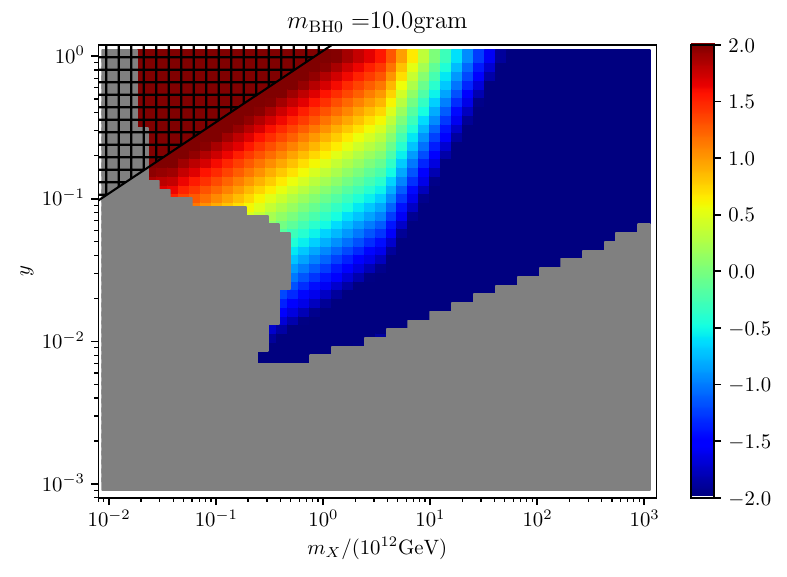}\includegraphics[width=0.49\textwidth]{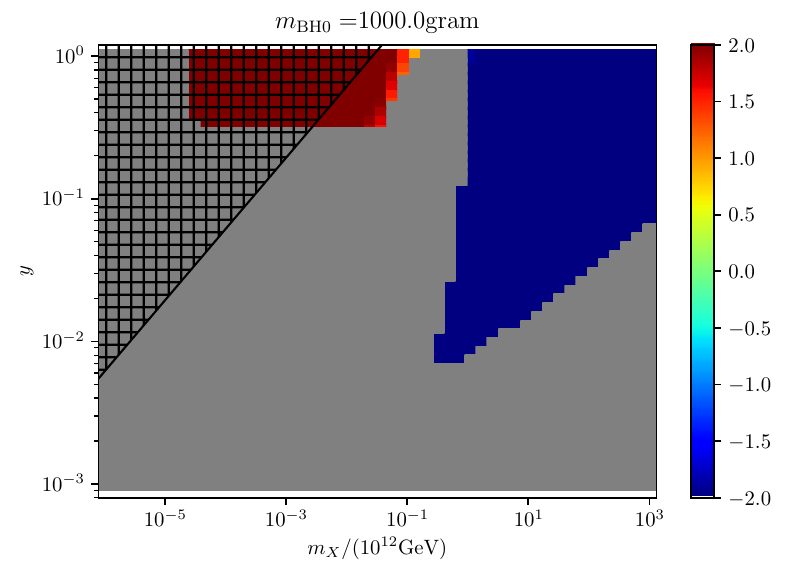}
	
	\caption{The viable parameter space in the $y$-$m_{X}$ plane. The gray and
		colored regions lead to insufficient and sufficient production of
		the $q_{e}$ asymmetry, respectively.  The color bars indicate the
		value of $\log_{10}(\Delta_{2}/\Delta_{1})$ where $\Delta_{1,2}$
		denote the TH and PBH contributions to the asymmetry. 
		In the hatched region, the produced asymmetry is expected to be washed out by two-to-two scattering---see Eq.~\eqref{eq:y-bound}.
		\label{fig:para}}
	
	\includegraphics[width=0.49\textwidth]{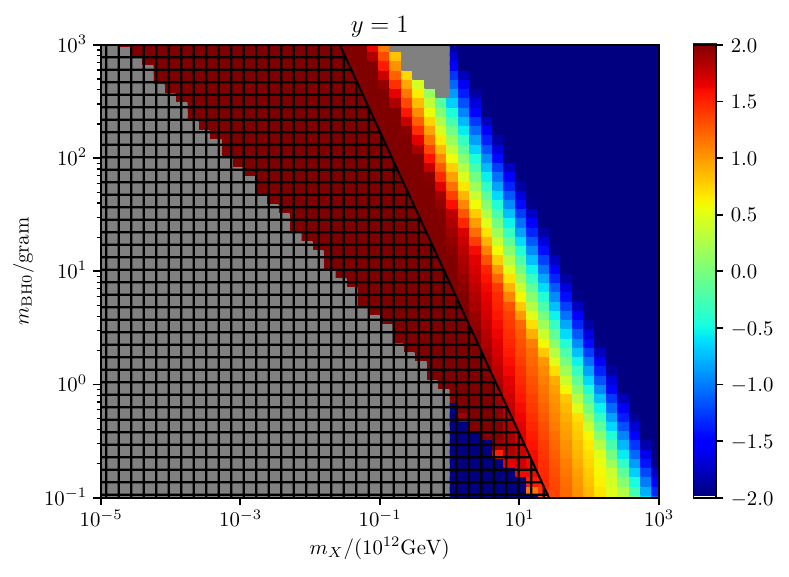}\includegraphics[width=0.49\textwidth]{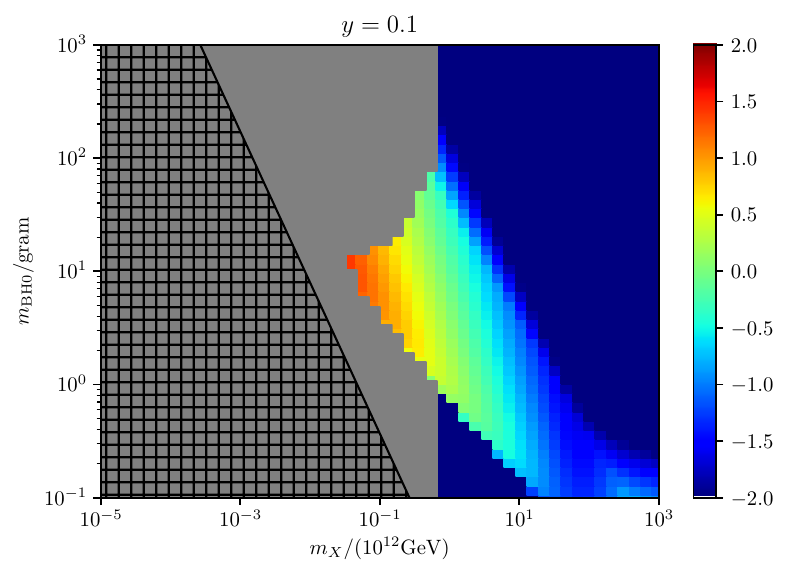}
	
	\caption{The viable parameter space in the $m_{{\rm BH0}}$-$m_{X}$ plane.
		The color coding is the same as Fig.~\ref{fig:para}. \label{fig:para-2}}
\end{figure*}

Our analytical approach in Sec.~\ref{sec:ana} allows us to understand
the evolution of the PBH-produced asymmetron $X$ qualitatively, and
to estimate the resulting asymmetry accurately within certain ranges.
Nevertheless, a numerical approach based on solving Boltzmann equations
is desirable for more accurate and comprehensive calculations. 

The Boltzmann equations governing the evolution of $X$ and the electron
asymmetry read
\begin{align}
\frac{dY_{1}}{da} & =\frac{\Gamma_{X}r_{1}}{Ha}\left[Y^{{\rm eq}}-Y_{1}\right],\label{eq:-19}\\
\frac{dY_{2}}{da} & =\frac{1}{Ha}\left[a^{3}n_{\text{BH}}\Gamma_{\text{BH}\to X}-Y_{2}\Gamma_{X}r_{2}\right],\label{eq:-20}\\
\frac{d\Delta_{1}}{da} & =\frac{\Gamma_{X}}{Ha}\left[\epsilon r_{1}\left(Y_{1}-Y^{{\rm eq}}\right)-\frac{r_{1}}{2}\frac{n_{X}^{\text{eq}}}{n_{e_{R}}^{\text{eq}}}\Delta_{1}\right],\label{eq:-21}\\
\frac{d\Delta_{2}}{da} & =\frac{\Gamma_{X}}{Ha}\left[\epsilon r_{2}Y_{2}-\frac{r_{1}}{2}\frac{n_{X}^{\text{eq}}}{n_{e_{R}}^{\text{eq}}}\Delta_{2}\right],\label{eq:-22}
\end{align}
where $Y\equiv n_{X}a^{3}$ and $\Delta\equiv q_{e}a^{3}$ with $a$
the scale factor. The subscripts 1 and 2 denoting the thermal and
PBH contributions, respectively. The superscript ``eq'' indicates
equilibrium values. The relativistic factors $r_{1,2}\equiv\langle m_{X}/E\rangle_{1,2}$
account for the time dilution of relativistic $X$ decay.  

The last terms in Eqs.~\eqref{eq:-21} and \eqref{eq:-22} are the
so-called washout terms. They are proportional to $\Delta_{1,2}$
and impose exponential suppression on the produced asymmetry until
the coefficient $n_{X}^{\text{eq}}/n_{e_{R}}^{\text{eq}}$ becomes
negligibly small. 

Equations \eqref{eq:-19} and \eqref{eq:-21} can be solved analytically
by making a few approximations in the strong and weak washout regimes
and combing the results via analytical interpolation~\cite{Buchmuller:2004nz}.
The result is
\begin{equation}
\frac{\Delta_{1}}{n_{\gamma}a^{3}}=\frac{3}{8}\epsilon\kappa_{f},\label{eq:-25}
\end{equation}
where $\kappa_{f}$ is a complicated function of $K\equiv\Gamma_{X}/H|_{T\to m_{X}}$.
Its full expression is given in Appendix \ref{sec:kappa}. In the
weak washout regime ($K\ll1$), $\kappa_{f}$ reduces to  $9\pi^{2}K^{2}/64$.
 Equations \eqref{eq:-25} with the full expression of $\kappa_{f}$
can provide a rather accurate estimate of the thermal contribution
to $q_{e}$---see the black dashed lines in the lower panels of Fig.~\ref{fig:boltzmann}. 

Numerically solving the Boltzmann equations is straightforward. 
Figure \ref{fig:boltzmann} shows the numerical solutions for three
examples representing three different regimes, as discussed below. 

\vspace{1ex}
\noindent $\blacksquare$ (i): strong washout with late PBH evaporation.

In the left panels, we set $m_{{\rm BH0}}=1$ gram, $\beta=10^{-4}$,
$m_{X}=10^{13}$ GeV, $y=1$ and $\epsilon=y^{2}/(8\pi)$. The $K$
value of this example is large, $K\approx1.4\times10^{3}\gg1$. As
is shown in the plot, the large coupling $y$ leads to strong washout
within  ${\cal O}(0.1)\lesssim m_{X}/T\lesssim{\cal O}(10)$.\footnote{The sharp drop at $m_{X}/T\approx0.4$ is caused by $q_{e}$ crossing
zero.} The thermal production of the asymmetry is washed out rapidly after
$X$ enters equilibrium and before $n_{X}^{{\rm eq}}$ is exponentially
suppressed, while the PBH production of the asymmetry can avoid this
strong washout if the evaporation lasts sufficiently long. 

\vspace{1ex}
\noindent $\blacksquare$ (ii): strong washout with early PBH evaporation.

For comparison, in the middle panels we show another example with
$m_{{\rm BH0}}=1$ gram, $\beta=10^{-3}$, $m_{X}=10^{11}$ GeV, $y=1$
and $\epsilon=y^{2}/(8\pi)$. Due to the smaller $m_{X}$, the PBHs
fully evaporate within the strong washout regime. Hence the PBH production
of the asymmetry is almost entirely washed out. 

\vspace{1ex}
\noindent $\blacksquare$ (iii): weak washout.

If the coupling is sufficiently weak, $X$ cannot reach thermal equilibrium
at temperatures relevant for the production of asymmetry. This is
in the weak washout regime. In the right panels, we plot an example
with $m_{{\rm BH0}}=1$ gram, $\beta=10^{-4}$, $m_{X}=10^{13}$ GeV,
 $y=10^{-3}$ and $\epsilon=y^{2}/(8\pi)$. The $K$ value of this
example is small, $K\approx3.5\times10^{-4}\ll1$, leading to only
 weak washout of the thermal and PBH contributions to the asymmetry.
As is shown in the lower right panel, the asymmetry $|q_{e}|$ produced
from PBHs is many orders of magnitude higher than the thermal one.
This is a noteworthy difference, given that the thermal and PBH productions
of $n_{X}$ shown in the upper panel are roughly comparable. The reason
for this substantial difference is that the PBH curve for $n_{X}$
deviates more from the equilibrium curve than the thermal one.

From the three examples, we can see that the PBH lifetime is crucial
for the asymmetry produced from PBHs to evade strong washout and to
dominate over the thermal production. Note, however, that their lifetime
cannot be too long if PBHs are employed in wash-in leptogenesis, which
requires that some SM yukawa interactions are not equilibrated. For
wash-in leptogenesis via $|q_{e}|$, it requires that the PBHs should
evaporate before the electron yukawa interaction reaches equilibrium,
i.e. $T_{{\rm eva}}\gtrsim100$ TeV, corresponding to $m_{\text{BH0}}\lesssim4.6$
kg.

Recently there have been several studies pointing out the existence of locally hot plasma around a PBH caused by the Hawking radiation heating up the local region~\cite{Das:2021wei,He:2022wwy, Hamaide:2023ayu}. This might lead to an additional washout effect for this work, if the emitted asymmetrons decay in the local region around a PBH. The calculation of this effect would involve the thermal profile and the dynamic evolution of the local region, which could be investigated in future work.   

\section{Full scan of the parameter space\label{sec:Full-scan}}

To identify the viable parameter space, we numerically solve the Boltzmann
equations for $m_{\text{BH0}}\in[0.1,\ 10^{3}]$ gram, $m_{X}\in[10^{7},\ 10^{16}]$
GeV, and $y\in[10^{-3},\ 1]$. This allows us to perform a scan of
the parameter space. The results are presented in Figs.~\ref{fig:para}
and \ref{fig:para-2}, where the colored/gray regions produce sufficient/insufficient
baryon asymmetry. In addition, we also impose the condition that allows the two-to-two scattering to be neglected [see Eq.~\eqref{eq:y-bound}] on the figures as hatched regions.

 The upper bound of $m_{{\rm BH}0}$ is set at $10^{3}$ gram  according
to the discussion  at the end of Sec.~\ref{sec:boltzmann}. The lower
bound  is set at $0.1$ gram because the formation of lighter PBHs
would require $T_{0}\gtrsim10^{16}$ GeV, which would be above the
energy scale of inflation constrained by the Planck 2018 data~\cite{Planck:2018jri}.

The asymmetry produced by PBHs is proportional to $\beta$ and $\epsilon$.
For $\beta$, we set $\beta=2\times10^{-5}\times(1\ {\rm g}/m_{{\rm BH0}})$
so that the universe is always radiation dominated during PBH evaporation.
Larger $\beta$ is possible but the evolution involves PBH domination---see
e.g. \cite{Bernal:2022pue}. If one further increases $\beta$ to
values above $1.1\times10^{-6}\times(10^{4}\ \text{g}/m_{{\rm BH0}})^{17/24}$,
it would lead to an overly large contribution to the cosmological
$N_{{\rm eff}}$ parameter via gravitational waves~\cite{Papanikolaou:2020qtd,Domenech:2020ssp}.
As for $\epsilon$, we only require that it is below $y^{2}/(8\pi)$.
In practice, we set $\epsilon$ at this upper limit and compute the
resulting asymmetry. If it is above the required value, then we consider
it as a viable sample point, since the produced asymmetry can be readjusted
to the desired value by reducing $\epsilon$.

In Fig.~\ref{fig:para} we present the parameter space in the the
$y$-$m_{X}$ plane with $m_{{\rm BH0}}$ fixed at $0.1$, $1$, $10$
and $1000$ gram. In these plots, there are some red or yellow regions
in which the asymmetry is dominantly produced by PBH. Typically these
regions are in the regime of strong washout with late PBH evaporation,
corresponding to regime (i) discussed in Sec.~\ref{sec:boltzmann}.
However, for $m_{{\rm BH0}}=0.1$ gram, there is a small yellow region
around $y\sim2\times10^{-2}$ and $m_{X}\sim3\times10^{14}$ GeV.
This regime actually corresponds to the weak washout regime. Note
that as $m_{{\rm BH0}}$ increases, the viable parameter space for
weak washout quickly vanishes, while the viable parameter space in
the strong washout regime remains robust. 

In Fig.~\ref{fig:para-2}, we fix $y$ at given values and scan over
$m_{X}$ and $m_{{\rm BH0}}$.  For $y=1$,  as is shown on the left panel,  
the width of the red region generally increases as $m_{\rm BH0}$ increases, 
implying that later PBH evaporation generally leads to more effective production of the asymmetry.
The ratio of PBH to thermal contributions $\Delta_2/\Delta_1$ increases for smaller $m_X$, which however is bounded from below by Eq.~\eqref{eq:y-bound}. 
When
$y$ is below certain values, as is illustrated by the right panel with $y=0.1$, the region favored by the regime of
strong washout with late PBH evaporation vanishes quickly. This is caused by the
suppression of the asymmetric decay parameter $\epsilon$ which is
proportional to $y^{2}$. 

\section{Summary and Conclusions \label{sec:con}}

Wash-in leptogenesis offers a novel mechanism for generating the baryon
asymmetry of the universe by utilizing possible primordial charge
asymmetries not necessarily limited to the commonly considered cases
of $B$, $L$ or $B-L$ charges.  The primordial charge asymmetries
required for wash-in leptogenesis can be readily produced by evaporating
primordial black holes (PBHs). 

To demonstrate the viability of this idea, we consider a rather simple
scenario, in which PBHs emit heavy particles (dubbed as the asymmetrons)
that can asymmetrically decay to electrons and thus generate the primordial
electron asymmetry, $q_{e}$. If the asymmetrons  are emitted by
PBHs at a relatively late epoch of the universe, the generated asymmetry
can effectively evade strong wash-out, as illustrated by Fig.~\ref{fig:bench}
for a specific benchmark. 

By solving the Boltzmann equations with both PBH and thermal contributions
taken into account, we find that there are three very different regimes:
(i) strong washout with late PBH evaporation, (ii) strong washout
with early PBH evaporation, and (iii) weak washout. For each regime,
we demonstrate an example in Fig.~\ref{fig:boltzmann}. Our results
reveal that the PBH lifetime is crucial for the asymmetry produced
from PBHs to evade strong washout and to dominate over the thermal
production. 

We further conduct  a full scan of the parameter space (see Figs.~\ref{fig:para}
and \ref{fig:para-2}) and find that strong washout with late PBH
evaporation is the most favored regime. The former only requires that
the coupling of the asymmetron with the electron is not suppressed,
while the latter requires relatively large PBH masses, corresponding
to longer lifetimes.  For instance, an ${\cal O}(0.1\sim 1)$ asymmetron
coupling combined with an initial mass of $10^{3}$ g for the PBHs
can easily lead to sufficient primordial asymmetry before wash-in
leptogenesis, if the asymmetron mass is within $[10^{9},\ 10^{11}]$
GeV. 

In conclusion, our research demonstrated that PBHs in joint with wash-in
leptogenesis can act as a very generic and efficient ``asymmetry-producing
machine'', since any asymmetry injected by PBHs into the SM thermal
bath at a sufficiently early epoch would eventually be reprocessed
by wash-in leptogenesis  and converted to the baryon asymmetry. Our
novel approach features invulnerability to strong washout and the
capability to employ more generic $CP$ violating sources.  This opens
new avenues for exploring the origin of the baryon asymmetry.  


\begin{acknowledgments}
The authors would like to thank the organizers of the CERN Neutrino Platform Pheno Week in
March 2023, where this project was initiated. The work of K.\,S.\ is supported by the
Deutsche Forschungsgemeinschaft (DFG) through the Research Training Group, GRK 2149: Strong
and Weak Interactions\,---\,from Hadrons to Dark Matter.
X.-J.\,X.\ is supported in part by the National Natural Science Foundation
of China under grant No.~12141501 and also supported by CAS Project
for Young Scientists in Basic Research (YSBR-099). 
\end{acknowledgments}

\appendix
\begin{widetext}

\section{Basic formulae for PBH evaporation \label{sec:Basic-PBH}}

Schwarzschild PBHs radiate particles at the rate given by Eq.~\eqref{eq:-39}.
The graybody factor $v_{i}$ in Eq.~\eqref{eq:-39} is suppressed
for $T_{{\rm BH}}\ll m_{i}$ where $m_{i}$ is the mass of the particle
emitted. For $T_{{\rm BH}}\gg m_{i}$, it behaves like blackbody radiation.
The actual dependence of the graybody factor on the energy and the
mass can be quite complicated (see e.g. \cite{Cheek:2021odj,Masina:2021zpu,Auffinger:2020afu}
for recent discussions). In this work, we approximate the graybody
factor as follows~\cite{Bernal:2022pue}:
\begin{equation}
v_{i}=\frac{27E^{2}}{64\pi^{2}T_{{\rm BH}}^{2}}\cdot\Theta\thinspace,\label{eq:-23}
\end{equation}
where
\begin{equation}
\Theta\equiv\begin{cases}
1 & m_{i}/T_{{\rm BH}}<x_{{\rm cri.}}\\
0 & m_{i}/T_{{\rm BH}}>x_{{\rm cri.}}
\end{cases}\thinspace,\ x_{{\rm cri.}}\approx3.6\thinspace.\label{eq:-24}
\end{equation}
With the graybody factor in Eq.~\eqref{eq:-23}, we can compute the
energy loss rate of a PBH:
\begin{equation}
\frac{dm_{{\rm BH}}}{dt}=-\sum_{i}\int\frac{d^{2}N_{i}}{dtdE}EdE=g_{{\rm BH}}\frac{m_{{\rm pl}}^{4}}{m_{{\rm BH}}^{2}}\thinspace,\label{eq:-3-1}
\end{equation}
where $g_{{\rm BH}}\approx7\times10^{-5}g_{\star}$ and $g_{\star}$
counts all bosonic ($\times1$) and fermionic ($\times7/8$) degrees
of freedom that are lighter than $T_{{\rm BH}}$. The PBHs considered
in this work typically have $T_{\text{BH}}\gg$TeV so all SM particles
should be included as massless degrees of freedom. Assuming the dominance
of the SM contribution to $g_{{\rm BH}}$, we can treat $g_{{\rm BH}}$
as a constant: $g_{{\rm BH}}\approx7.5\times10^{-3}$. 

For constant $g_{{\rm BH}}$, Eq.~\eqref{eq:-3-1} has the following
analytical solution:
\begin{equation}
m_{{\rm BH}}=m_{{\rm BH}0}\left(1-3g_{{\rm BH}}m_{\text{pl}}^{4}\frac{t-t_{0}}{m_{{\rm BH}0}^{3}}\right)^{1/3}.\label{eq:-4}
\end{equation}
where $t_{0}$ and $m_{{\rm BH}0}=m_{{\rm BH}}(t_{0})$ denote the
initial time and mass of the PBH. Eq.~\eqref{eq:-4} implies that
the PBH mass vanishes at $t-t_{0}=\tau_{{\rm BH}}$ with
\begin{equation}
\tau_{{\rm BH}}=\frac{m_{{\rm BH}0}^{3}}{3g_{{\rm BH}}m_{\text{pl}}^{4}}\thinspace,\label{eq:-5}
\end{equation}
which is the lifetime of the PBH. 

The emission rate of $X$ from a PBH, according to Eqs.~\eqref{eq:-39}
and \eqref{eq:-23}, is given by
\begin{equation}
\Gamma_{\text{BH}\to X}\equiv\int\frac{d^{2}N_{X}}{dtdE}dE\approx\frac{81\zeta(3)}{256\pi^{3}}T_{{\rm BH}}\cdot\Theta\approx0.012T_{{\rm BH}}\cdot\Theta\thinspace.\label{eq:-6-1}
\end{equation}

If $X$ decays rapidly after the emission, then the energy spectrum
is proportional to Eq.~\eqref{eq:-39}. 

If  $X$ is sufficiently long-lived, then the energy spectrum is
computed by
\begin{equation}
\frac{dN_{X}}{dE}=\int\frac{d^{2}N_{X}}{dtdE}dt=\int_{T_{{\rm BH0}}}^{\infty}\frac{d^{2}N_{X}}{dtdE}\frac{8\pi m_{{\rm BH}}^{4}}{g_{\text{BH}}m_{\text{pl}}^{6}}dT_{{\rm BH}}\thinspace.\label{eq:-36}
\end{equation}
Assuming the Boltzmann statistics and $T_{{\rm BH}}\gtrsim m_{X}/x_{{\rm cri.}}$,
we obtain 
\begin{equation}
\frac{dN_{X}}{dE}=\frac{81m_{\text{pl}}^{2}\left[1-e^{-x}\left(1+x+\frac{x^{2}}{2}+\frac{x^{3}}{6}+\frac{x^{4}}{24}\right)\right]}{8192\pi^{6}g_{\text{BH}}T_{{\rm BH0}}^{3}x^{3}}\thinspace,x\equiv E/T_{{\rm BH0}}\thinspace.\label{eq:-34}
\end{equation}
Note that $1+x+\frac{x^{2}}{2}+\frac{x^{3}}{6}+\frac{x^{4}}{24}=e^{x}+{\cal O}(x^{5})$
so at small $x$, Eq.~\eqref{eq:-34} is proportional to $x^{5}e^{-x}/x^{3}\approx x^{2}e^{-x}$.
At $x\gg1$, Eq.~\eqref{eq:-34} implies
\begin{equation}
\frac{dN_{X}}{dE}\approx\frac{81m_{\text{pl}}^{2}}{8192\pi^{6}g_{\text{BH}}T_{{\rm BH0}}^{3}x^{3}}\thinspace,(\text{for}\ x\gg1)\thinspace.\label{eq:-35}
\end{equation}
Therefore,  unlike usual thermal distributions which is suppressed
by $e^{-x}$ at large $x$, Eq.~\eqref{eq:-34} features a very ``hard''
spectrum at high energies. 

In Sec.~\ref{sec:boltzmann}, we introduced the $r_{2}$ factor,
which is evaluated as follows:
\begin{equation}
r_{2}\equiv\left\langle \frac{m_{X}}{E}\right\rangle _{2}=\frac{\int\frac{m_{X}}{E}fd^{3}p}{\int fd^{3}p}=\begin{cases}
0.46\frac{m_{X}}{T_{{\rm BH}}} & \text{for short-lived }X\\
\frac{1}{3}\frac{m_{X}}{T_{{\rm BH}0}} & \text{for long-lived }X
\end{cases}.
\label{eq:PBH-r}
\end{equation}
Here $f$ takes $\frac{d^{2}N_{X}}{dtdE}$ in Eq.~\eqref{eq:-39}
for short-lived $X$, or $\frac{dN_{X}}{dE}$ in Eq.~\eqref{eq:-34}
for long-lived $X$.

\section{Analytical solutions of the Boltzmann equations\label{sec:kappa}}

The Boltzmann equations considered in this work can be written into
the following form
\begin{equation}
\frac{dY(a)}{da}=X(a)-F(a)Y(a)\thinspace,\label{eq:-26}
\end{equation}
where $X(a)$ and $F(a)$ denote some generic functions of $a$. It
is known that differential equations in the form of Eq.~\eqref{eq:-26}
have the following solution:
\begin{equation}
Y(a')=G(a')\int_{0}^{a'}\frac{X}{G}da\thinspace,\ \ G(a')\equiv e^{-\int_{0}^{a'}Fda}\thinspace.\label{eq:-27}
\end{equation}
In principle, Eq.~\eqref{eq:-27} can be used to analytically solve
the Boltzmann equations. In practice, one has to take various approximations
when applying it to specific processes. For the thermal production
of asymmetry considered in this work, we can adopt results from a
similar calculation in Ref.~\cite{Buchmuller:2004nz}---see Eqs.~(62), (63), (68), and (69)
therein.  The result is 
\begin{equation}
\frac{q_{e}}{n_{\gamma}}=\frac{3}{8}\epsilon\kappa_{f}\thinspace,\label{eq:-32}
\end{equation}
where the full expression of the $\kappa_{f}$ function is given by
\begin{equation}
\kappa_{f}(K)\approx2\frac{1-e^{-\frac{2}{3}z_{B}K\overline{N}}}{z_{B}K}-2e^{-\frac{2}{3}N}\left(e^{\frac{2}{3}\overline{N}}-1\right),\label{eq:-28}
\end{equation}
with
\begin{align}
z_{B} & \approx1+\frac{1}{2}\log\left[1+\frac{\pi K^{2}}{1024}\left(\log\frac{3125\pi K^{2}}{1024}\right)^{5}\right],\label{eq:-29}\\
N & =\frac{9\pi K}{16}\thinspace,\label{eq:-31}\\
\overline{N} & =\frac{N}{\left(1+\sqrt{4N/3}\right)^{2}}\thinspace.\label{eq:-31-1}
\end{align}
\end{widetext}

\bibliographystyle{JHEP}
\bibliography{ref}

\end{document}